\documentclass[letter,longauth]{aa}
\usepackage{natbib,twoopt}
\usepackage{amsmath}
\usepackage{wasysym}
\usepackage{multirow}
\usepackage{longtable}
\usepackage{graphicx}
\usepackage{mathtools}
\usepackage{blindtext}
\usepackage[varg]{txfonts}

\bibpunct{(}{)}{;}{a}{}{,} 

\newcommand{\gsc}{GSC~07396-00759}
\newcommand{\degree}{$^{\circ}$}
\newcommand{\bpmg}{$\beta$ Pic MG}

\usepackage{color}

\begin{document}

\title{New disk discovered with VLT/SPHERE around the M star \gsc \thanks{Based on data collected at the European Southern Observatory, Chile (ESO Program 198.C-0298)}}
\author{E. Sissa\inst{1},
J. Olofsson\inst{2,3,4},
A. Vigan\inst{5},
J.C. Augereau\inst{6},
V. D'Orazi\inst{1},
S. Desidera\inst{1},
R. Gratton\inst{1}, 
M. Langlois\inst{7,5},
E. Rigliaco\inst{1}, 
A. Boccaletti\inst{8}, 
Q. Kral\inst{8,9},
C. Lazzoni\inst{1,10},
D. Mesa\inst{1,11},
S. Messina\inst{12}, 
E. Sezestre\inst{6},
P. Th\'ebault\inst{8},
A. Zurlo\inst{13,14,1},
T. Bhowmik\inst{8},
M. Bonnefoy\inst{6}, 
G. Chauvin\inst{6,13}, 
M. Feldt\inst{2}, 
J. Hagelberg\inst{6}, 
A.-M. Lagrange\inst{6}, 
M. Janson\inst{15,2}
A.-L. Maire\inst{2}, 
F. M\'enard\inst{6},
J. Schlieder\inst{16,2},
T. Schmidt\inst{8}, 
J. Szul\'agyi \inst{17,18},
E. Stadler\inst{6},
D. Maurel\inst{6},
A. Delboulb\'e\inst{6},
P. Feautrier\inst{6},
J. Ramos\inst{2},
F. Rigal\inst{19}
}

\authorrunning{E. Sissa et al.}
\offprints{E. Sissa,  \\
   \email{elena.sissa@oapd.inaf.it} }
\institute{
 INAF-Osservatorio Astronomico di Padova,  Vicolo dell'Osservatorio 5, I-35122, Padova, Italy  
\and Max Planck Institut f\"ur Astronomie, K\"onigstuhl 17, 69117 Heidelberg, Germany
\and Instituto de F\'isica y Astronom\'ia, Facultad de Ciencias, Universidad de Valpara\'iso, Av. Gran Breta\~na 1111, Playa Ancha, Valpara\'iso, Chile
\and N\'ucleo Milenio Formaci\'on Planetaria - NPF, Universidad de Valpara\'iso, Av. Gran Breta\~na 1111, Valpara\'iso, Chile
\and Aix-Marseille Universit\'e, CNRS, LAM (Laboratoire d'Astrophysique de Marseille) UMR 7326, 13388, Marseille, France 
\and Universit\'{e} Grenoble Alpes, CNRS, IPAG, 38000 Grenoble, France 
\and CRAL, UMR 5574, CNRS, Universit\'e de Lyon, Ecole Normale Suprieure de Lyon, 46 Alle d'Italie, F-69364 Lyon Cedex 07, France
\and LESIA, Observatoire de Paris-Meudon, CNRS, Universit\'{e} Pierre et Marie Curie, Universit\'{e} Paris Diderot, 5 Place Jules Janssen, F-92195 Meudon, France 
\and Institute of Astronomy, University of Cambridge, Madingley Road,
Cambridge CB3 0HA, UK
\and Dipartimento di Fisica e Astronomia - Universita' di Padova, Vicolo dell'Osservatorio 3, I-35122, Padova, Italy   
\and INCT, Universidad De Atacama, calle Copayapu 485, Copiap\'{o}, Atacama, Chile
\and INAF-Osservatorio Astrofisico di Catania, Via S. Sofia 78, I-95123 Catania, Italy
\and Núcleo de Astronomía, Facultad de Ingeniería, Universidad Diego Portales, Av. Ejercito 441, Santiago, Chile
\and Departamento de Astronom\'ia, Universidad de Chile, Casilla 36-D, Santiago, Chile
\and Department of Astronomy, Stockholm University, AlbaNova University Center, 106 91 Stockholm, Sweden
\and Exoplanets and Stellar Astrophysics Laboratory, Code 667, NASA Goddard Space Flight Center, Greenbelt, MD 20770, USA
\and Institute for Particle Physics and Astrophysics, ETH Zurich,
Wolfgang-Pauli-Strasse 27, CH-8093 Zurich, Switzerland
\and Center for Theoretical Astrophysics and Cosmology, Institute for Computational Science, University of Zürich, Winterthurestrasse 190, CH-8057 Zürich, Switzerland
\and  Anton Pannekoek Institute for Astronomy, Science Park 904, NL-1098 XH Amsterdam, The Netherlands
}

\date{Received  /
Accepted }

\abstract{
Debris disks are usually detected through the infrared excess over the photospheric level of their host star. The most favorable stars for disk detection are those with spectral types between A and K, while the statistics for debris disks detected around low-mass M-type stars is very low, either because they are rare or because they are more difficult to detect. Terrestrial planets, on the other hand, may be common around M-type stars. Here, we report on the discovery of an extended (likely) debris disk around the M-dwarf \gsc. The star is a wide companion of the close accreting binary V4046 Sgr. The system probably is  a member of the $\beta$ Pictoris Moving Group. We resolve the disk in scattered light, exploiting high-contrast, high-resolution imagery with the two near-infrared subsystems of the  VLT/SPHERE instrument, operating in the YJ bands and the H2H3 doublet.  The disk is clearly detected up to 1.5\arcsec\ ($\sim110$ au) from the star and appears as a ring, with an inclination $i\sim83$\degree, and a peak density position at $\sim 70$\,au. The spatial extension of the disk suggests that the dust dynamics is affected by a strong stellar wind, showing similarities with the AU Mic system that has also
been resolved with SPHERE. The images show faint asymmetric structures at the widest separation in the northwest side. We also set an upper limit for the presence of giant planets to  $2 M_J$. Finally, we note that the 2 resolved disks around M-type stars of 30 such
stars observed with SPHERE are viewed close to edge-on, suggesting that a significant population of debris disks around M dwarfs could remain undetected because of an unfavorable orientation.
} 



\maketitle

\section{Introduction}
Circumstellar disks are commonly detected around young stars both by ground- and space-based telescopes in a wide portion of the spectrum \citep[see, e.g.,][]{moro2013, krivov2010, matthews2014}, from scattered light and near-infrared wavelengths with facilities such as the Hubble Space Telescope or \emph{Spitzer}, to mid- and far-infrared wavelengths with facilities such as \emph{Herschel}, and up to submillimeter wavelengths for radio-telescopes such
as ALMA. 
Primordial circumstellar disks around M dwarfs have been shown to typically have longer lifetimes than those around more massive stars \citep{carpenter2006,luhman2012}. However, many surveys found that their evolved counterparts, the debris disks, are quite rare \citep{plavchan2005, lestrade2006, gautier2007, lestrade2009, avenhaus2010}.  
Up to now, very few low-mass M-type stars older than 15-20\,Myr show excesses, especially in the near- and mid-infrared, and even fewer dwarfs have resolved images of disks: the AU Mic disk was first detected in scattered light with high-contrast imaging \citep{kalas2004}, whereas  GJ 581 \citep{lestrade2012},  and possibly Proxima Cen \citep{anglada2017,macgregor2018} were imaged in direct thermal emission with \emph{Herschel} and ALMA, respectively\footnote{To this short list of direct imaging resolved debris disks, we can also add  the two debris disks of TWA 7 \citep[Olofsson et al., submitted]{choquet2016} and TWA 25 \citep{choquet2016}. These debris disks were found around stars younger than 15\,Myr and belong to an age bin where the frequency of primordial disks is still significant.}.  Several of these disks have peculiar features, such as multiple rings, spirals \citep[][Olofsson et al. submitted.]{anglada2017}, and fast-moving arc-like structures \citep{boccaletti2015,boccaletti2018}. \\
The paucity of debris disk detections around low-mass M-type stars may have two main reasons: disks around such stars are in fact  less common and/or they are more difficult to detect.
On one hand, the formation of planetesimals may be strongly inhibited both by external photoevaporation due to intense far-UV radiation field that typically influences these objects \citep{adams2004} and to close stellar flybys \citep{lestrade2011}. On the other hand, M-type stars are very cool objects and thus they are less luminous than stars of earlier type. This implies that dust particles are less heated and faintly emit at long wavelengths, making these excesses hard to detect with current instrumentation.\\
 Debris disks around young M stars are also relevant to explore  the possible link between their occurrence and the presence of terrestrial planets in the system \citep{raymond2011}, which are very frequent around low-mass stars \citep{bonfils13, dressing2015}. The presence of planetary systems around both GJ 581 and Proxima \citep{mayor2009,anglada2016}  is very promising in this perspective. \\
 For these reasons, the detection of spatially resolved  disks around low-mass M-type stars is of high interest.
We present the newly imaged \gsc\ likely debris disk with the
instrument SPHERE \citep[Spectro-Polarimetric High-contrast Exoplanet REsearch;][]{beuzit2008}.
\gsc\ is a very active but not accreting M1-type star associated with V4046 Sgr \citep{sacy, kastner2011}, which is itself a close binary with accretion signature \citep{stempels2004} and a gas-rich circumbinary disk \citep{rosenfeld2013}. The two systems are
probably members of the $\beta$ Pictoris Moving Group  \citep[\bpmg, $24\pm3$\,Myr, ][]{sacy,malo2014}. More details on the stellar properties can be found in Appendix~\ref{sec:param}. 

\section{Observations and results}
\label{sec:obs}
\begin{table}
        \caption{Overview of observational SPHERE data sets for \gsc. }            
        \label{tab:data} 
        \centering
\begin{tabular}{lc}          
                \hline\hline
                Parameter & Value\\
                \hline                       
                date & 2017 Jun 15 \\
                filters & YJ H2H3 \\
                 total integration time  [s] & 6016\\
                 total rotation of the field of view  [\degree]& 112.82  \\
                  $\tau_0$ [ms] & 20 \\
                  SR              & 0.73 \\
                \hline                                             
        \end{tabular}
\end{table}
\begin{figure*}
\centering
\includegraphics[width=\textwidth]{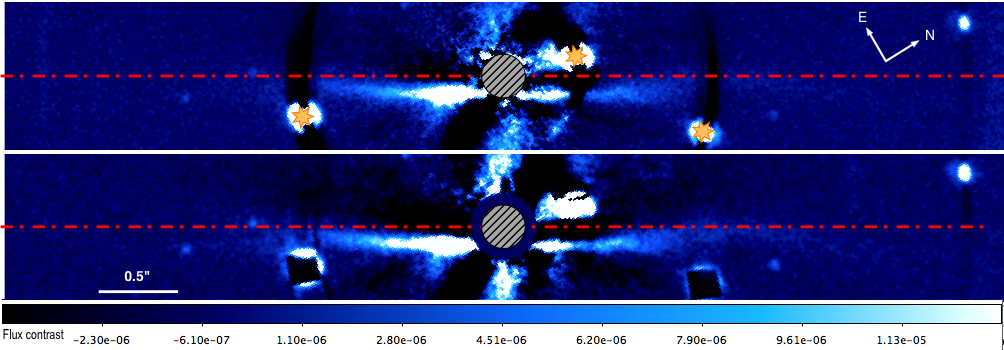} \\
\caption{New disk detected around \gsc. IRDIS 5 modes PCA of H2H3 images rescaled to stellar flux before (top) and after (bottom) the subtraction of the most contaminant stars,  identified with star-like points in the upper figure. The gray-shaded region in the center corresponds to the SPHERE coronagraph. The dash-dotted line is parallel to the disk semi-major axis and crosses \gsc. The bright background objects are masked behind star-shaped regions in the upper image.}
\label{IRDIS_best}
\end{figure*}
We observed \gsc\ during the SPHERE Guaranteed Time Observation as part of  the SpHere INfrared survey for Exoplanets (SHINE), in the night of June 15, 2017. The sky conditions were very good, with an average coherence time longer than 10 ms and an average Strehl ratio of $\sim$65\%, which is excellent considering the faintness of the source \citep[$R=12.01$,][]{ucac5}. 
This target was observed in the standard IRDIFS mode, using IFS \citep{claudi2008}  in the YJ bands (between 0.95 and 1.35 $\mathrm{\mu m}$, 1.7\arcsec$\times$1.7\arcsec\ field of view; FoV) and IRDIS \citep{dohlen2008} in dual-band imaging H2H3 mode \citep[at 1.59 and 1.67 $\mathrm{\mu m}$;][11\arcsec$\times$11\arcsec\ FoV]{vigan2010}  simultaneously (see Tab. \ref{tab:data}).
The data were reduced with version 0.15.0 of the SPHERE Data Reduction and Handling (DRH) pipeline \citep{pavlov2008}, and the images were further processed  using the SpeCal software (Galicher et al. in prep.).  Both packages are hosted at the SPHERE Data Center (DC) in Grenoble\footnote{\url{http://sphere.osug.fr/spip.php?rubrique16&lang=en}}\citep{sphereDC}. Additional details on the adopted procedures are provided in \cite{zurlo2014}, \cite{mesa2015}, and \cite{maire2016astrom}.

\subsection{Point source detection and planetary upper limits}
The IFS YJ wavelength-collapsed image gives a $5\sigma$ contrast limit  $<$11 mag in  the innermost regions ($<0.25$\arcsec), which drops to $\sim12$ mag for separations wider than $\sim 0.3$\arcsec. The IRDIS images lead to a contrast limit of less than 12\,mag at a separation of $\sim 0.6$\arcsec\ and $\sim13.5$\,mag at separations larger than 1.3\arcsec. These performances led to the detection of 109 candidate companions (see Appendix \ref{sec:CCs} for their characterization) and a previously unknown faint disk (Fig. \ref{IRDIS_best}). None of the candidate companions are probably physically linked to \gsc, as expected given its projected position with respect to the galactic plane.
Furthermore, we converted the contrast limits into upper limits
on the mass of possible unseen companions. Using the theoretical atmospheric models AMES-COND \citep{allard2003} with age and distance as in Table \ref{t:param}, we  obtained an upper limit for undetectable companions of 4 M$_J$ in  the innermost regions ($<0.25$\arcsec), which drops to 2 M$_J$ for separations wider than $\sim 0.6$\arcsec (see Fig. \ref{fig:contrast} for more details).
\subsection{Spatially resolved disk}
\begin{figure}
\includegraphics[width=1.1\columnwidth, trim={-1cm 0 0 0}, clip]{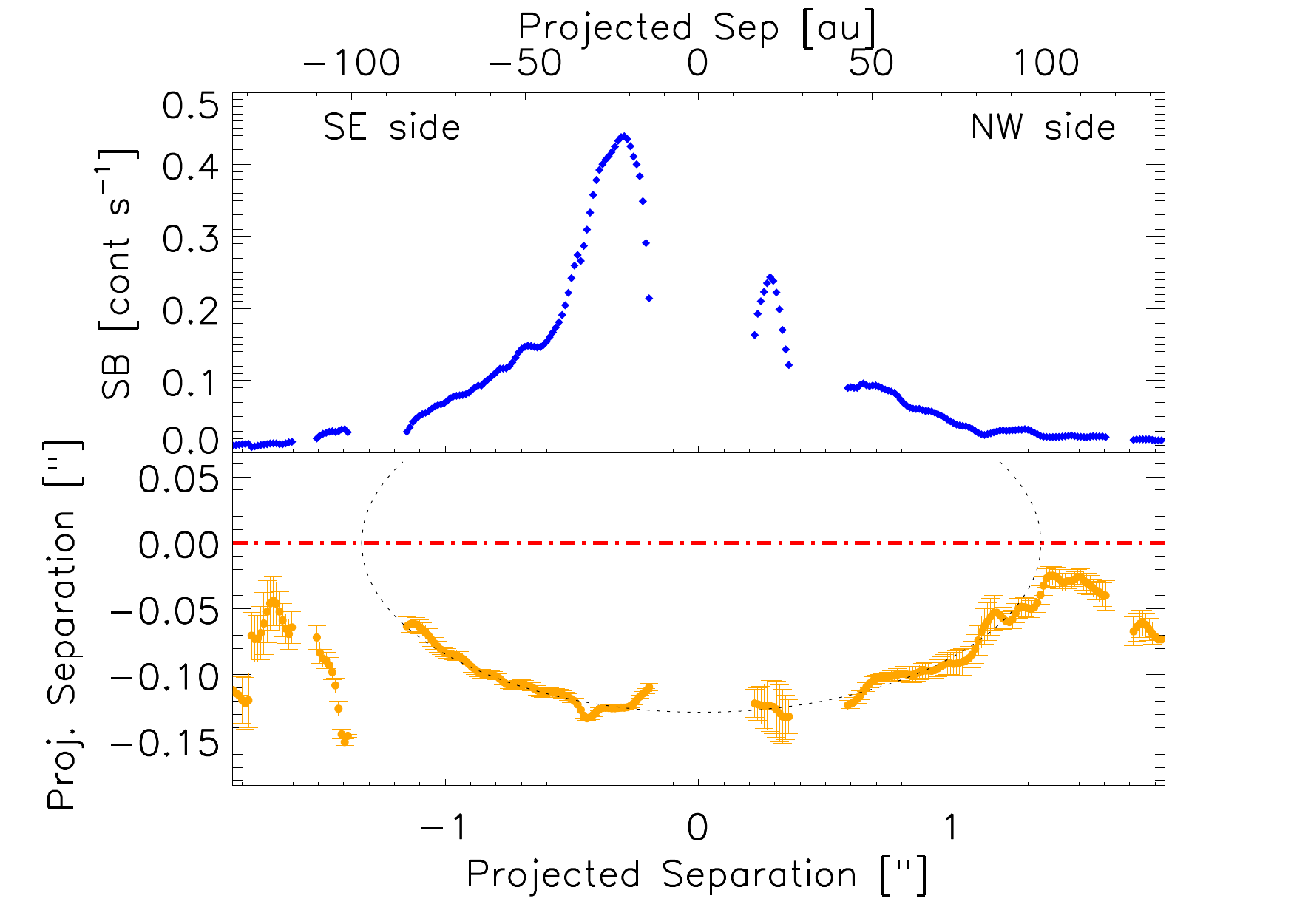}
\caption{Lower panel: Spine plot of the disk after background star removal on the five-mode PCA IRDIS image. Upper panel: Surface brightness profile along the disk as defined by the spine, in arbitrary units.} 
\label{fig:spine}
\end{figure}
We resolved a highly inclined disk that is probably coplanar with the stellar rotation (see Appendix~\ref{sec:param}) and extends more than 1.5\arcsec\  from the central star. The disk appears brighter in the southeast side (left part in Fig.~\ref{IRDIS_best}). In Fig.~\ref{fig:spine} we plot the position of the maximum brightness (referred to as the disk spine in the following) with respect to the apparent semi-major axis (obtained with disk forward-modeling, see Sec.~\ref{sec:model}) as a function of the separation from the star, together with its intensity. This is obtained by extracting slices from the image, perpendicular to the apparent semi-major axis, and by fitting a Gaussian profile to derive the projected separation and the intensity. However,  the presence of at least three very bright stars alters the light distribution of the disk after applying the aggressive algorithms of SpeCal, as shown in Fig.~\ref{IRDIS_best}. Thus, we first proceeded to remove
the stars, as described in Appendix~\ref{sec:star_removal}, and we obtained an image less affected by starlight residuals and self-subtraction effects.  Fig.~\ref{fig:spine} confirms that the SE side is noticeably brighter than the NW side.  Assuming that the disk is not flared, a ring of material with radius $r=1.34 \pm   0.01$\arcsec\ and inclination $i=84.5 \pm 3.6$\degree (dotted line) can properly describe the observed spine up to 1.2\arcsec.  Moreover, in the outer part of both sides of the disk, the light distribution indicates some swept-back material, a warp of the disk, as clearly detected with the spine plot.
\section{Forward-modeling of the SPHERE/IRDIS data}
\label{sec:model}
Because of the self-subtraction effects related to the angular differential imaging process (e.g., \citealp{Milli2012}), one cannot directly compare a disk model image to the reduced images. An alternative possibility is to perform forward-modeling and inject negative models in the datacube before performing the angular differential process, and to try to remove any signal arising from the circumstellar disk.
The modeling strategy is described in Appendix \ref{sec:mod_strategy}.

\begin{table}
\centering
\caption{Best-fit results for the disk modeling of the SPHERE observations.\label{tab:results}}
\begin{tabular}{@{}lcc@{}}
\hline\hline
Parameter               & Uniform prior   & Best-fit value \\
\hline
$r_0$ [au]              & $[50, 80]$      & $69.9_{-0.8}^{+0.9}$ \\
$i$ [$^{\circ}$]        & $[75, 88]$      & $82.7_{-0.1}^{+0.1}$ \\
$\alpha_{\mathrm{in}}$  & $[1.5, 10]$     & $2.8_{-0.2}^{+0.2}$ \\
$\alpha_{\mathrm{out}}$ & $[-10, -1.5]$   & $-2.6_{-0.1}^{+0.1}$ \\
$\phi$ [$^{\circ}$]     & $[140, 160]$    & $148.9_{-0.1}^{+0.1}$ \\
$g$                     & $[0, 0.99]$     & $0.50_{-0.01}^{+0.01}$ \\
$f$                     & $[7, 9]$        & $7.76_{-0.01}^{+0.01}$ \\
\hline
\end{tabular}
\tablefoot{We list the reference radius $r_0$, where the dust density distribution peaks, the inner and the outer slopes of the dust density distribution $\alpha_{\mathrm{in}}$ and $\alpha_{\mathrm{out}}$, the inclination $i$ and the position angle $\phi$, the parameter $g$ that governs the scattering efficiency, and the scaling factor.}
\end{table}

\begin{figure*}
\centering
\includegraphics[width=\hsize]{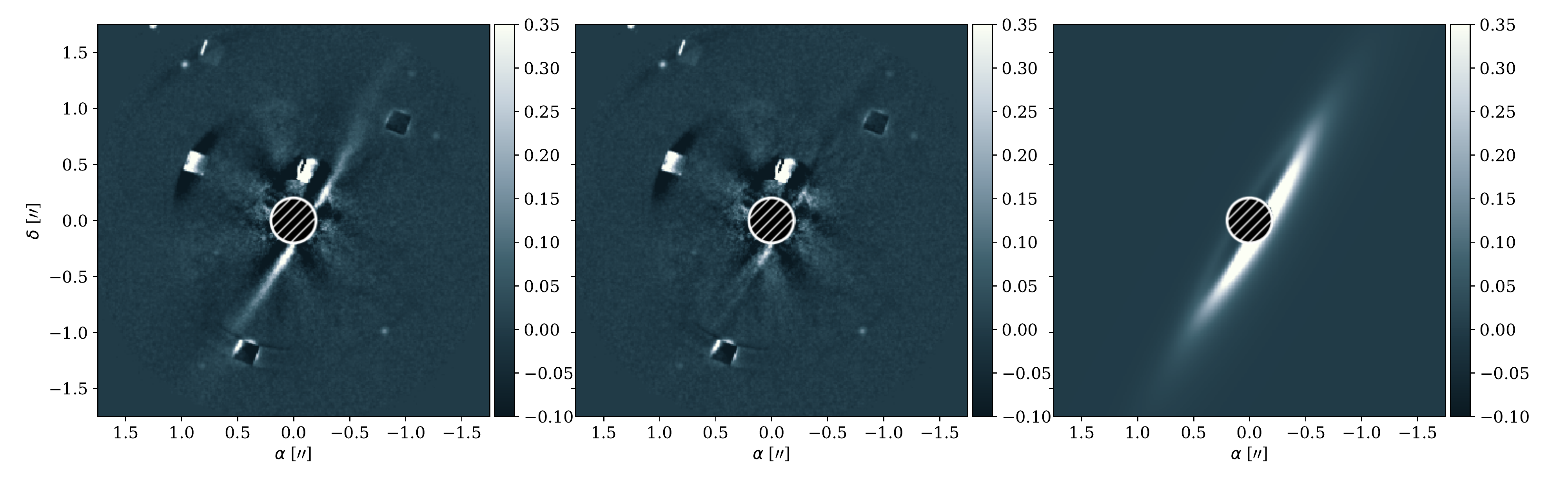}
\caption{{\it Left to right}: IRDIS data, residuals, and best-fit model. All images have the same linear scaling. Stars were removed as described in Appendix~\ref{sec:star_removal}. North is up, east is left. The central shaded area corresponds to the numerical mask. Color bars are in units of counts per second.}
\label{fig:results}
\end{figure*}

We determined the most probable values for each of the free parameters  (the reference radius $r_0$, where the dust density distribution peaks, the inner and the outer slopes of the dust density distribution $\alpha_{\mathrm{in}}$ and $\alpha_{\mathrm{out}}$, the inclination $i$ and the position angle $\phi$, the parameter $g$ that governs the scattering efficiency, and the scaling factor  $f$, see Appendix \ref{sec:mod_strategy}) from the probability density function. The one- and two-dimensional density distributions are shown in Fig.~\ref{fig:corner} (made using the \texttt{corner} package, \citealp{corner}). From the distributions we also derived the $68$\% confidence intervals that are reported in Table\,\ref{tab:results}. The best-fit model, along with the observations and the residuals, is presented in Figure\,\ref{fig:results} with the same linear scaling. Most of the signal coming from the disk has been cancelled: there are no significant residuals toward the NW side of the star, while larger amplitude residuals are left on the SE side, close to the numerical mask. This would further suggest that the disk shows some level of asymmetry, as discussed in Sec.~\ref{sec:obs}.

\begin{figure}
\centering
\includegraphics[width=0.35\textwidth, trim={0 0 0 1cm}, clip]{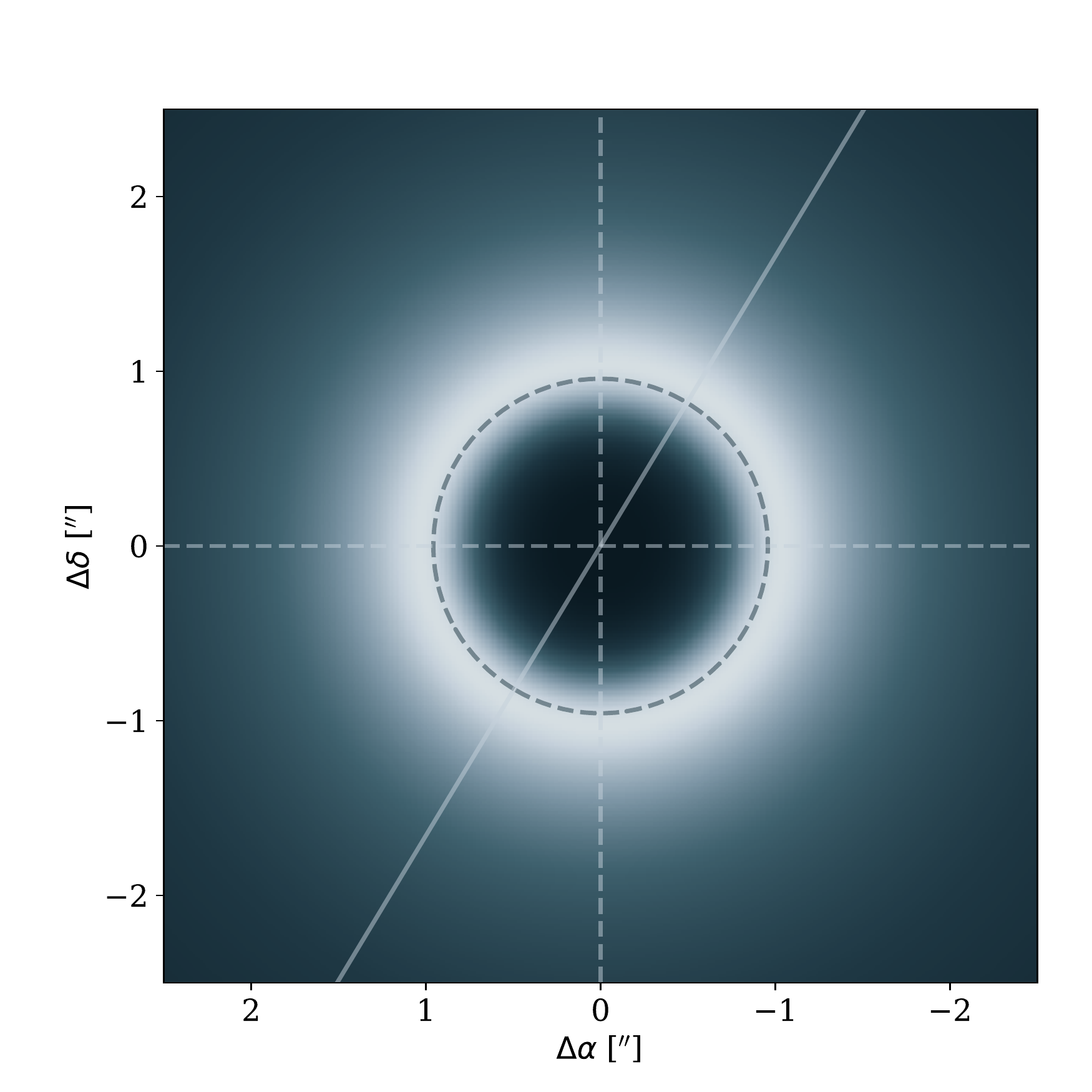}
\caption{Cross-section weighted density plot of the best-fitting model as seen from face-on. The circle marks the location of the reference radius $r_0$ and the diagonal marks the position angle of the disk.  The image has a linear scaling.}
\label{fig:density}
\end{figure}

We found that the disk is highly inclined ($i\sim83^{\circ}$), with a position angle $\phi\sim149^{\circ}$ that it is relatively extended with a peak for the dust density distribution at $\sim70$\,au, and inner and outer slopes of $2.8$ and $-2.6$, respectively (the assumed radial profile is described in Appendix~\ref{sec:mod_strategy}). For the scattering phase function, we found a coefficient $g\sim0.5$. Because of the high inclination, the disk is unresolved  along the semi-minor axis, and we were unable to place further constraints on the entire phase function (e.g., \citealp{Milli2017}). To illustrate how extended the disk is according to our modeling results, Fig.~\ref{fig:density} shows the cross-section weighted density distribution of the best-fit model, as seen from face-on. 

\begin{figure}
\centering
\includegraphics[width=\columnwidth]{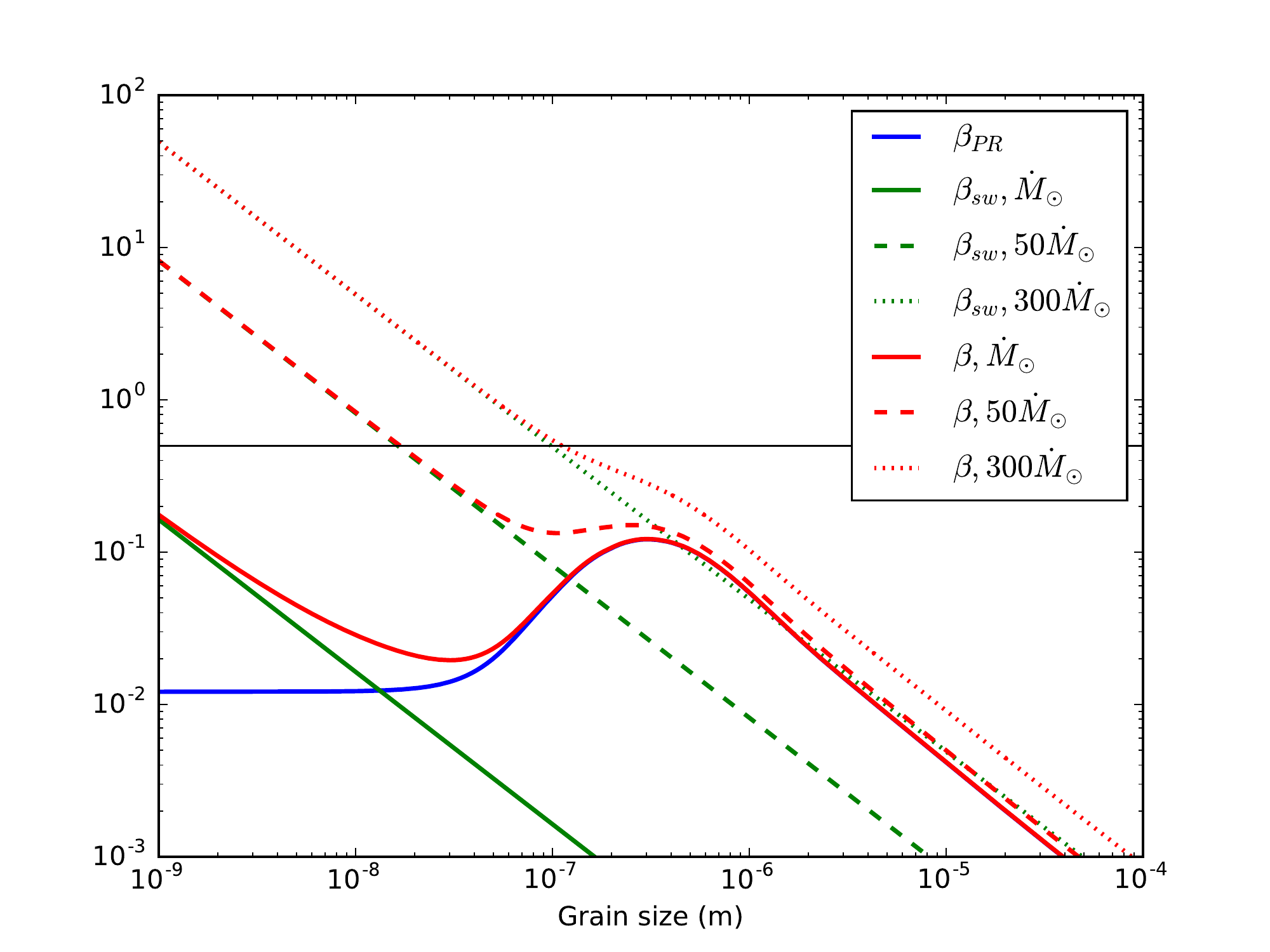} 
\caption{$\beta$ as a function of grain size. The material used for $\beta_{PR}$ is M1 of \cite{schueppler2015}. Three hypotheses for the mass-loss rate of the star are shown (solid, dashed, and dotted lines). The horizontal solid black line is the upper limit for bound trajectories assuming zero eccentricity for the parent body, see \cite{sezestre2017} for further details.}
\label{fig:beta}
\end{figure}
Interestingly, the surface density beyond $r_0$ seems to settle around a $\alpha_{\mathrm{out}}+1\sim-1.5$ profile that is expected for dust grains subject to a strong radial pressure force like the stellar radiation pressure, and produced by collisional grinding of larger bodies arranged in a possibly narrow “birth” ring \citep{Strubbe2006,Thebault2008,kral2013}. Because the luminosity of \gsc\ is low, the radiation pressure force never overcomes the gravitational force, as shown in Fig.~\ref{fig:beta} (blue curve), where $\beta_{PR}$ is the ratio between the two forces. This strongly suggests that the radial extent of the debris disk could rather be the consequence of a strong stellar wind pressure force, as discussed in \citet{Augereau2006} for AU Mic. This would be consistent with the star being young and active (Appendix~\ref{sec:param}). The wind pressure force can be parametrized by the ratio $\beta_{SW}$ between the wind pressure and gravitational forces. The net pressure force acting on a grain is then defined by $\beta=\beta_\mathrm{PR}+\beta_\mathrm{SW}$ \citep{sezestre2017}. Our preliminary results shown in Fig.~\ref{fig:beta} suggest that the requirement of having a strong enough pressure force ($\beta > 0.5$) is obtained for submicron-sized grains and sufficiently high stellar mass loss rates (at least a few tens of the solar mass-loss rate).

Additionally, we compared our best-fitting model with the observed spectral energy distribution (SED). The photometry was gathered using the VO Sed Analyzer tool (VOSA\footnote{\url{http://svo2.cab.inta-csic.es/theory/vosa/index.php}}, \citealp{Bayo2008}), and no excess is detected, with the farthest wavelength at which the system has been observed being the \textit{WISE} $22$\,$\mu$m point ($<$4.49 mJy, \citealt{cutri2014}). Overall, our modeling results are compatible with a non-detection of the disk at wavelengths shortward of $22$\,$\mu$m. However, given the lack of far-IR photometric points, we cannot constrain the dust mass from the SED, therefore we incrementally increased it until we matched the \textit{WISE}/W4 upper limit for non-detection, and found M$_{\mathrm{dust}} \sim 0.33$\,M$_{\oplus}$. This would correspond to a fractional luminosity  upper limit with respect to the stellar luminosity $L_{\mathrm{disk}} / L_\star \sim 4.3 \times 10^{-3}$, which is still compatible with some of the brightest debris disks, such as HR~4796A.

\begin{figure}
\centering
\includegraphics[width=\hsize]{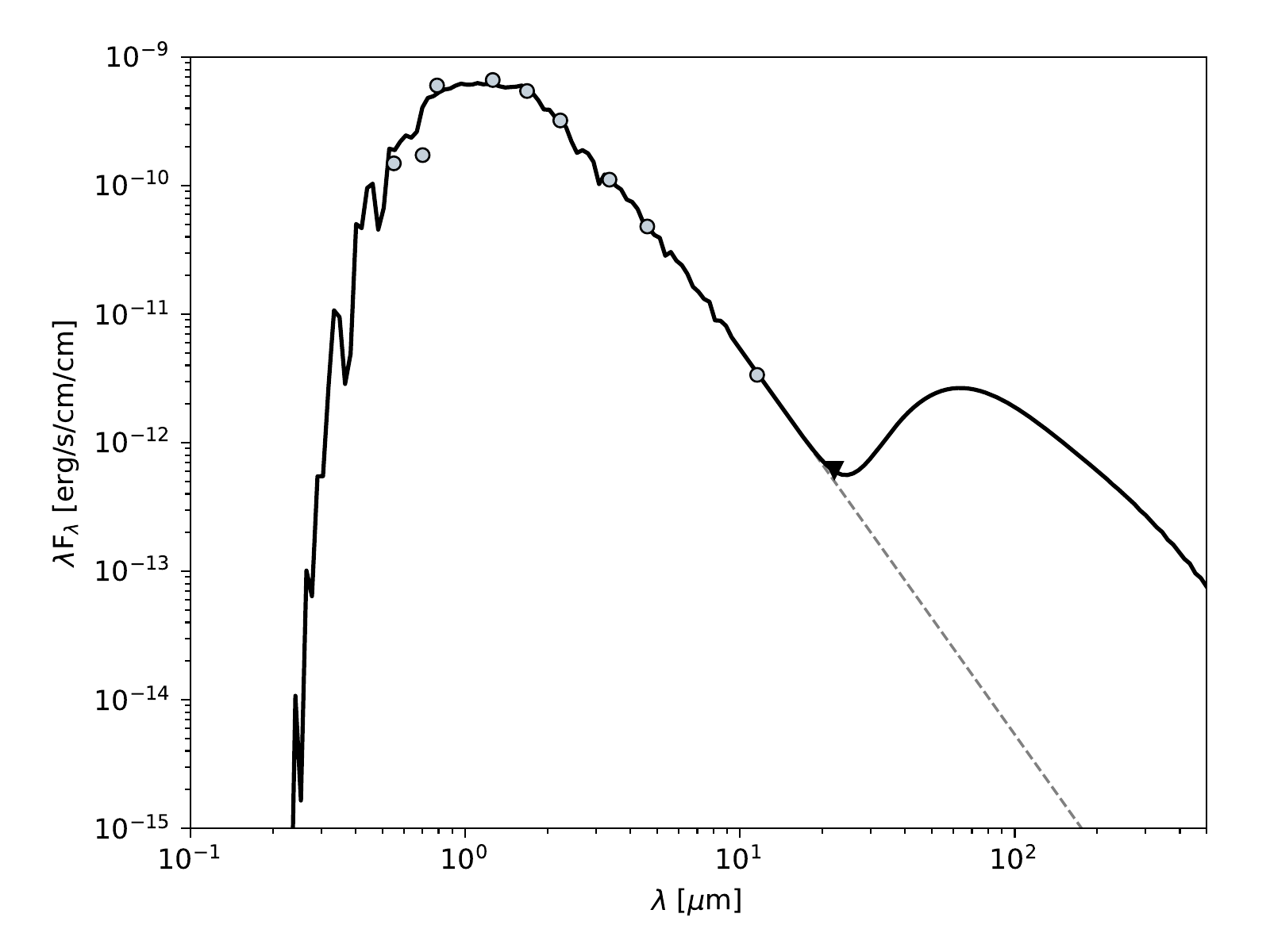}
\caption{SED of GSC\,07396-00759, the photometry is shown in light blue, the stellar model in dashed gray, and the model with the maximum possible IR-excess compatible with observations is plotted in solid black.}
\label{fig:SED}
\end{figure}

\section{Discussion}
\label{sec:discussion}
\subsection{Comparison of the disks of \gsc\ and V4046 Sgr}
The association with V4046 Sgr, which is still accreting and still surrounded by a gas-rich disk, raises the question of the nature of the disk around \gsc. While none of the available observational constraints allows us to unambiguously settle this issue, several properties of this disk combined point toward an evolved optically thin debris disk instead of a primordial gas-rich disk. First, the central star does not show evidence of accreting gas. Then, there is the non-detection of an IR excess, which indicates that the disk is cold and suggests a central cavity. Moreover, for a nearly edge-on configuration, an optically thick disk should cause some absorption of the light coming from the star, which we do not observe. In addition, our modeling indicates a fractional luminosity upper limit of $4.3\times10^{-3}$ that is a typical value for debris disks regime and points toward an optically thin system, since primordial or even transition disks like HD~141569 have fractional luminosities upward of $\sim 0.01$ \citep{wyatt2015}. Finally, the surface brightness slope $\sim r^{-1.5}$ in the outer regions corresponds to the signature expected for the halo of small dust particles that naturally forms, in the presence of stellar radiation pressure of wind, beyond a debris ring that collisionally produces these small grains \citep{Strubbe2006,Thebault2008}. \\
In this respect, the discovery of this probable debris disk is particularly interesting considering that the wide companion binary system V4046 Sgr instead has a gas-rich disk \citep[see, e.g., ][D'Orazi et al., in prep]{oberg2011,rapson2015, rapson2015b}. %
This makes the V4046 Sgr system an interesting laboratory for studying the different evolutionary timescales of coeval disks. \gsc\ is classified as a weak-line T-Tauri star \citep{kastner2011}, while the close binary V4046 Sgr  is still in accretion phase \citep{stempels2004}. Circumbinary disks around close binaries have been demonstrated to have longer lifetimes than disks around single stars \citep{alexander2012}. This might be due to the torques produced by the close binary onto the disk, which open a cavity in the disk and suppress the accretion, resulting in a longer disk lifetime. The analysis of this aspect will be further investigated in a forthcoming paper.   
We have indications (from the $v \sin i$ of the star) that the \gsc\ disk is highly inclined with respect to the rotation axis of the star, while the V4046 Sgr disk is coplanar with the binary orbit. Moreover, the two disks are not coplanar ($i=$33.5\degree\ and $PA=76$\degree\ for V4046 Sgr, \citealt{rosenfeld2013}).\\
The \gsc\ disk in scattered light appears as extended as the AU Mic disk, the most extended debris disk known around M dwarfs. Moreover,  the swept-back faintest part of the disk, visible at wider separations, resembles the Moth disk \citep[see, e.g., ][]{schneider2014,Olofsson2016}, while ripples in the spine have
previously been observed in the disk of AU Mic \citep{boccaletti2015,boccaletti2018}. The origin of these structures and their possible relation with the wide companion V4046 Sgr is beyond the scope of this work and will be studied with future observations optimized for disk characterization.
\subsection{Debris disk around M stars}
In the SHINE survey, we observed 30 M stars until September 2017. Twenty-four of them have high-quality observations obtained in good weather conditions.
Twelve objects belong to the \bpmg\ or the TW Hya MG and are therefore younger than 25 Myr. Only two disks were detected in the standard SHINE mode: AU Mic \citep{boccaletti2015} and \gsc.
TWA 7 was also observed in the SHINE survey, but its dusty rings were detected only in polarimetry, while no structures were observed around Proxima Cen in the standard SHINE mode \citep{mesa2017}.
 As discussed in \cite{Milli2012}, three main factors favor the detection of edge-on compared to face-on disks. Two of these factors are related to physical effects: i) forward-scattering is more efficient than scattering at angles close to 90\degree, and ii) in optically thin disks, the emitting column is longer for higher inclinations. The third factor is due to our image-analyzing procedure: the ADI process removes the azimuthal component and tends to cancel out circular structures, so that the final contrast for edge-on disks is $\sim$100 times brighter (5 mag) than the same disk observed face-on \citep{Milli2012}. 
Assuming uniform distribution of disk inclinations around M stars, the probability of having a disk between 82.7\degree\ and 90\degree\ is 12.7\%. Assuming SPHERE can detect all these disks, the probability of detecting at least two disks out of 12 observations of stars surrounded by a disk is 46\%. Assuming that only 50\% of M dwarfs host a disk, the probability of observing two disks out of 12 M dwarfs is 17\%; it becomes barely 1\% if 10\% of the young M dwarfs host a debris disk.
We conclude that either there is a preferential direction in the disk orientation in both the \bpmg\ and the TW Hya association (which is unexpected given the results by \citealt{menard2004}), or it is more likely that this kind of disk phase is a frequent stage in the evolution of M dwarfs.

\section{Summary}
\label{sec:conclusion}
A new disk around the M dwarf \gsc, probably a \bpmg\ member, was detected with SPHERE in the near-infrared. We can summarize our results as follows:
\begin{itemize}
\item the disk is nearly edge-on ($i=83$\degree) and extended ($r_0=70$ au, $\alpha_{\mathrm{in}}=2.3$ and $\alpha_{\mathrm{out}}=-2.7$);
\item the surface density slope in the outer regions is $\sim-1.5$;
\item the disk is asymmetric and shows swept-back wings at separations of about 1.2\arcsec\ and ripples in the spine of the disk on either side;
\item the disk probably consists of submicron-sized grains that
are affected by strong stellar winds as a result of high stellar mass loss rates;
\item the disk shows a low fractional luminosity ($ L_{\mathrm{disk}} / L_\star \leqslant 4 \times 
10^{-3}$) ; 
\item the stellar rotation appears to be coplanar with the disk;
\item the star is not accreting gas.
\end{itemize}
Even if there is no "smoking gun" proof, the system characteristics
all together tend to favor an evolved/debris disk nature for \gsc\ over a primordial/gas-rich disk. If confirmed, this is a very interesting discovery since this star and V4046 Sgr form a coeval physically bound system that would consist of a gas-rich circumbinary disk and a debris disk. Therefore it is of paramount importance to search for gas in the \gsc\ disk.\\ 
We detected the \gsc\ disk in the context of the SHINE survey, and these observations are therefore optimized for planet detection. The detection of two edge-on disks among the 24 M dwarfs observed so far indicates that this type of debris disk might be common among young M stars.
To further constrain disk properties, especially the most faint structure we detect in the NW side and a possible origin of its asymmetries, broad-band observations and polarimetric observations with IRDIS are needed. Moreover, observations with other instruments at  wavelengths longward of 22 $\mu$m will lead to a better estimation of the dust properties.

\begin{acknowledgements}
SPHERE is an instrument designed and built by a consortium consisting of IPAG (Grenoble, France), MPIA  (Heidelberg, Germany), LAM (Marseille, France), LESIA (Paris, France), Laboratoire  Lagrange (Nice, France), INAF Osservatorio Astronomico di Padova (Italy),  Observatoire de Gen\`{e}ve (Switzerland), ETH Zurich (Switzerland), NOVA (Netherlands), ONERA (France) and ASTRON (Netherlands) in collaboration with ESO. SPHERE was funded by ESO, with additional contributions from CNRS (France), MPIA (Germany), INAF (Italy), FINES (Switzerland) and NOVA (Netherlands). SPHERE also received funding from the European Commission Sixth and Seventh Framework Programmes as part of the Optical Infrared Coordination Network for Astronomy (OPTICON) under grant number RII3-Ct-2004-001566 for FP6 (2004-2008), grant number 226604 for FP7 (2009-2012) and grant number 312430 for FP7 (2013-2016). SPHERE Data Center is supported by the LabEx OSUG@2020, Investissement d’avenir - ANR10 LABX56.
The authors thank the ESO Paranal Staff for support for conducting the observations. E.S., R.G., D.M., S.D. and V.D. acknowledge support from the "Progetti Premiali" funding scheme of the Italian Ministry of Education, University, and Research. This work has been supported by the project PRIN-INAF 2016 The Cradle of Life - GENESIS-
SKA (General Conditions in Early Planetary Systems for the rise of life with SKA). J.~O. acknowledges support from ICM N\'ucleo Milenio de Formaci\'on Planetaria, NPF, and from the Universidad de Valpara\'iso. J.-C.A. acknowledges support from the "Programme National de Plan\'{e}tologie" (PNP) of CNRS/INSU co-funded by the CNES. E.R. is supported by the European Union's Horizon 2020 research and innovation programme under the Marie Sk\l odowska-Curie grant agreement No 664931. Q. K. acknowledges funding from the STFC via the Institute of Astronomy, Cambridge Consolidated Grant. F. M. acknowledges funding from ANR of France under contract number ANR-16-CE31-0013. We thank P. Delorme and E. Lagadec (SPHERE Data Centre) for their efficient help during the data reduction process. E.S. thanks I. Carleo and G. Munaretto for the technical support in data analysis. This research makes use of VOSA, developed under the Spanish Virtual Observatory project supported from the Spanish MICINN through grant AyA2011-24052. 
\end{acknowledgements}

\bibliographystyle{aa} 
\bibliography{gsc7396} 

\begin{appendix}
\section{Stellar properties}
\label{sec:param}
\begin{center}
\begin{table}
\caption{Stellar parameters of \gsc.}
\label{t:param}
\begin{tabular}{lcl}
\hline\hline
Parameter      & Value  & Ref \\
\hline
V (mag)                       &  12.78  & \citet{messina2017}  \\
B$-$V (mag)                   &   1.36  & \citet{messina2017}  \\
V$-$I (mag)                   &   2.14  & \citet{messina2017}  \\
J (mag)                       &  9.443$\pm$0.023  & 2MASS \\
H (mag)                       &  8.766$\pm$0.038  & 2MASS \\
K (mag)                       &  8.539$\pm$0.023  & 2MASS \\
Distance (pc)                 &   73.0  & \cite{sacy} \\
$\mu_{\alpha}$ (mas\,yr$^{-1}$)  &  3.3$\pm$1.1  & UCAC5 \\
$\mu_{\delta}$ (mas\,yr$^{-1}$)  &  -52.5$\pm$1.1  & UCAC5 \\
RV   (km\,s$^{-1}$)            &  -6.1$\pm$0.5  & this paper \\
U  (km\,s$^{-1}$)              &   -7.3 & this paper \\
V  (km\,s$^{-1}$)              &  -15.4 & this paper \\
W  (km\,s$^{-1}$)              &   -8.4 & this paper \\
ST                            &  M1IVe  & \citet{pecaut2013} \\
$T_{\rm eff}$ (K)               &  3630$\pm$50 & \citet{pecaut2013} \\
$L~(L_{\odot})$                &  0.14$\pm$0.02  & this paper \\
EW Li (m\AA)                  &   180$\pm$5 & this paper \\
EW H$_{\alpha}$               &   -1.696$\pm$0.155 & this paper \\
EW H$_{\beta}$                &   -1.672$\pm$0.319 & this paper \\
$v \sin i $  (km\,s$^{-1}$)    &   4.17$\pm$0.76 & this paper \\
$\log L_{X}/L_{bol} $  &  {-2.59} &  this paper \\
$P_{rot}$ (d)          &   12.05$\pm$0.50 & \citet{messina2017} \\
Age (Myr)             &  24$\pm$3  & \bpmg, \citet{bell2015} \\
$R_{star} (R_{\odot})$   &    0.95$\pm$0.08  & this paper \\
\hline
\end{tabular}
\end{table}
\end{center}
The source \gsc\ is a young  and very active M1 type star.
\cite{sacy} and \citet{kastner2011} proposed its association with
V4046 Sgr, itself a close binary with accretion signature and a prominent
circumbinary disk \citep{stempels2004,rosenfeld2013}. They are
probably members of the $\beta$ Pictoris Moving Group  \citep[\bpmg,][]{sacy,malo2014}, whose current age estimation is $24\pm3$\,Myr \citep{bell2015}.

The physical association between \gsc\ and V4046 Sgr
is strongly supported by the very similar proper motion of the
components from the UCAC5 proper motions catalog \citep{ucac5}, with differences at 1 mas/yr (compatible within the uncertainties).
We then assumed that V4046 Sgr and \gsc\ form
a wide, hierarchical triple system, and we adopted the distance
of V4046 Sgr \citep[73\,pc, ][]{sacy} for \gsc. The projected separation between the components is 12300\,au ($\sim 0.06$\,pc),  which is very wide but still compatible with a bound system, especially considering its young age \citep[see, e.g.,][]{andrews2017}.

The star has several indicators of youth, such as the weak lithium line,  coronal emission, and moderate rotation (see Table \ref{t:param}). 
A comparison with the \bpmg\ rotation versus color sequence indicates that 
rotation ($P=12.05$\,d) is slower than in the most the 
members, but a few similar outliers are observed, mostly stars with 
debris disks \citep{messina2017}. 
The lithium content is within the distribution of \bpmg\  members of similar colors, although a slightly higher lithium depletion could be expected considering the Li/rotation correlation observed in the  temperature range including our target \citep{messina2016}.

An effective temperature of 3632$\pm$19\,K was derived  
(from \citealp{pecaut2013}, weighted mean of their two determinations), while the M1 spectral type 
would correspond to 3630\,K in the scale Pecaut and Mamajek adopted for young stars. 
Hereafter we use 3630$\pm$50\,K.
By coupling this with the adopted magnitudes, the bolometric correction from
\citet{pecaut2013}, and the distance, we obtain a
luminosity of $0.14\pm0.02\,L_{\odot}$   and a stellar radius of 
$0.95\pm0.08\,R_{\odot}$. The position on the color-magnitude diagram is roughly consistent with the sequence of \bpmg\ for the adopted distance.

To further refine the stellar parameters, we considered the UVES \citep{uves} high-resolution spectra available in the ESO archive\footnote{Program ID. 088.C-0506(A) and 095.C-0437(B)}. We determined a lithium equivalent width EW(Li) of 180$\pm$5\,m\AA.  The $v\,\sin\,i$ was determined from spectral synthesis of isolated lines in the UVES spectrum, leading to 4.17$\pm$0.76\,km/s. This value,  combined with the stellar radius and rotation period from Table~\ref{t:param}, is compatible within $<1 \sigma$ with an edge-on stellar inclination. 

We also checked the presence of signatures of accreting gas. $\mathrm{H\alpha}$ and $\mathrm{H\beta}$ lines are detected in emission, and they present a double-peaked profile. A similar line profile is also observed in AU Mic, which is also a member of the \bpmg, has nearly the same stellar temperature \citep{pecaut2013}, and hosts a debris disk. \cite{houdebine1994} modeled these emission lines for AU Mic using a non-thermally heated chromosphere, finding a good match with the line profiles. Given the almost edge-on orientation of the disk, the $\mathrm{H\alpha}$ and $\mathrm{H\beta}$ line profiles might also be interpreted as emission coming from gas in Keplerian rotation around the star and not in any  accretion/ejection activity, however. If these activities were present, we would expect broader wings and more complex profiles in the hydrogen lines. 

There is no indication of close companions from radial velocities \citep[three RVs consistent within errors: ][and our own
determination on UVES spectra]{sacy,elliott2014} and previous imaging observations either \citep{galicher2016,janson2017}.

\section{Candidate companions}
\label{sec:CCs}
With only one SPHERE epoch for \gsc, the status of the 109 companion candidates detected in the IRDIS field of view could not be determined directly. Fortunately, this star was observed in the context of the International Deep Planet Survey \citep[IDPS;][]{Vigan2012,galicher2016} with the NIRC2 instrument at Keck. Deep observations were obtained in the $K^\prime$ filter on June 10,$^{}$ 2006, in pupil-stabilized mode (with 7.9\degree\ of field rotation) to perform angular differential imaging. These data were downloaded from the Keck archive and reanalyzed with the LAM-ADI pipeline \citep{Vigan2016} using a simple ADI procedure. The images were derotated to align north up and east left, and a spatial filter with a $5\times5\lambda/D$ kernel was applied to remove the smooth residual halo. The candidates visible in the field were then simply fitted with a two-dimensional Gaussian to identify their position with respect to the star. Taking into account the uncertainties on the centering of the stars and the uncertainty of the fit, we estimate a typical error of 0.5 pixel, that is, 4.98~mas in NIRC2, on the position of the candidates. For the estimation of the astrometry we did not consider any distortion or true-North correction. Finally, the candidates were manually cross-matched between the NIRC2 and IRDIS data.

\begin{figure}
    \centering
    \includegraphics[width=0.5\textwidth]{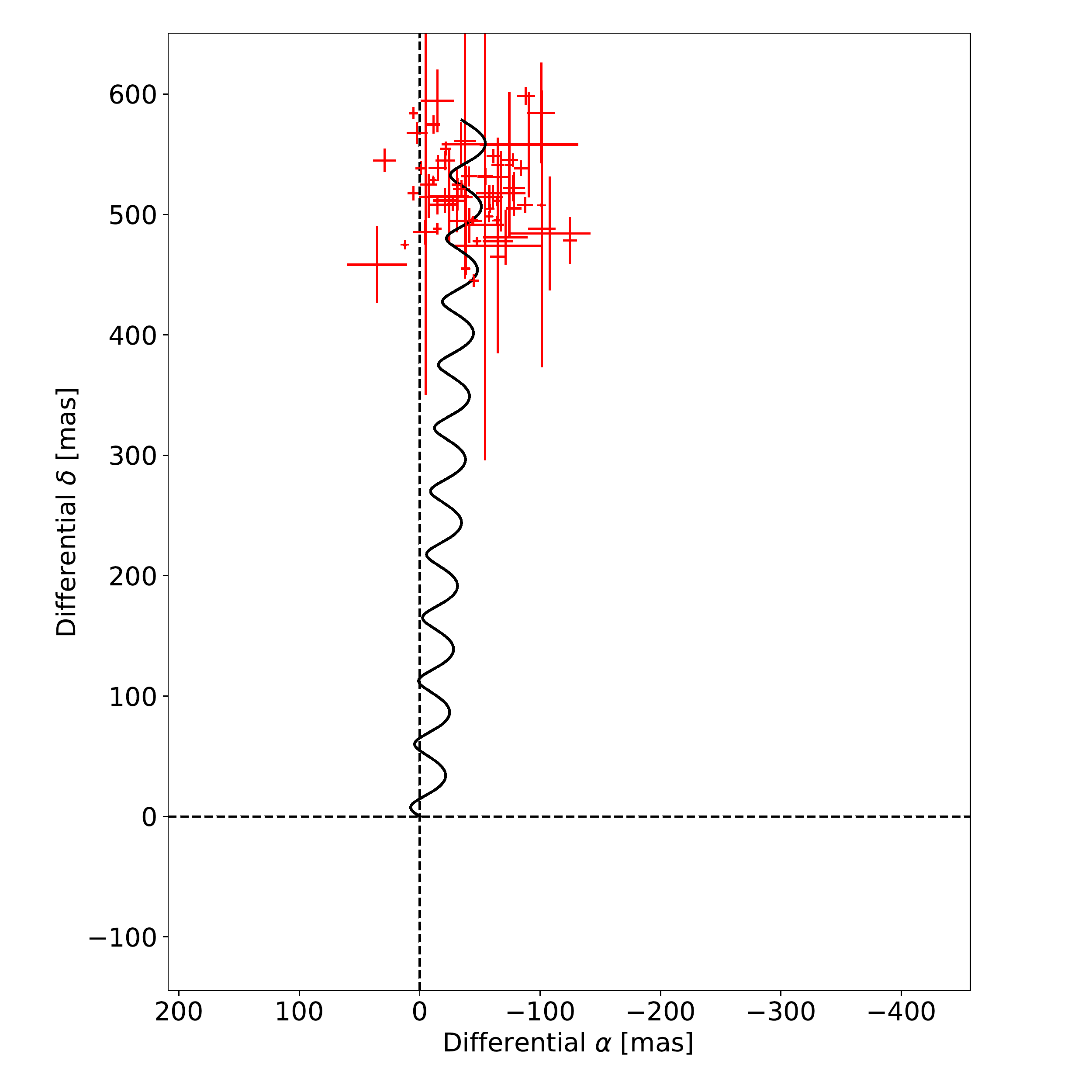}
    \caption{Differential right ascension and declination of the 70 candidates we cross-identified between the Keck/NIRC2 and SPHERE/IRDIS, compared to the track expected for a stationary background object. All 70 candidates are background stars.}
    \label{fig:all_cc}
\end{figure}

\begin{figure}
    \centering
    \includegraphics[width=0.5\textwidth]{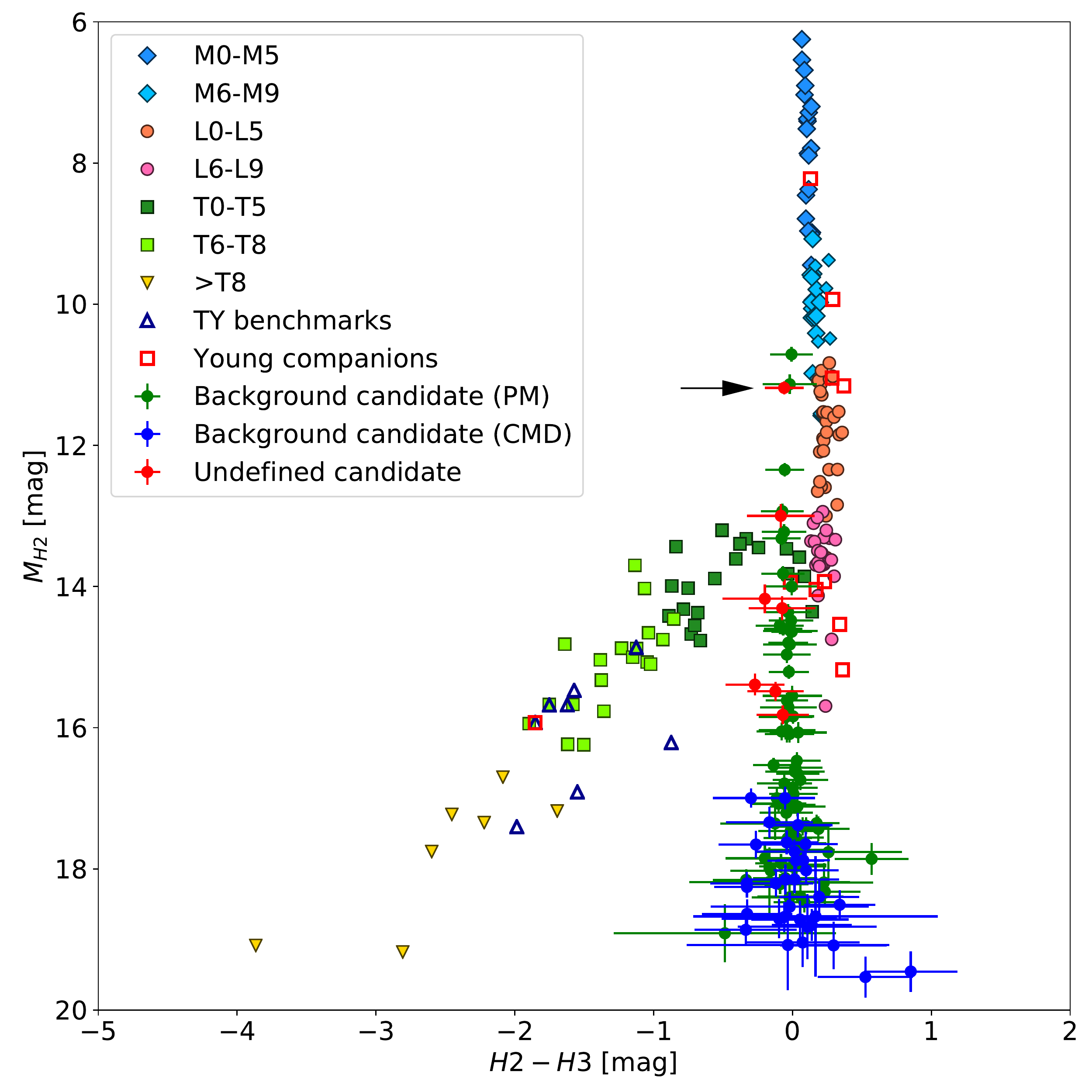}
    \caption{Color-magnitude diagram in the IRDIS H2 and H3 filters. The candidates detected in the IRDIS FoV are represented on top of the sequence of MLT objects from \citet{leggett2001}, \citet{Burgasser2014} and \citet{Schneider2015}, as well as additional young benchmark objects and companions. The candidates identified as background from proper motion (see Fig.~\ref{fig:all_cc}) are shown as green circles, those identified from the CMD as blue circles, and those for which the status remains undefined as red circles. The arrow marks the position of the candidate identified at a separation of 457~mas in the IFS FoV.}
    \label{fig:cmd}
\end{figure}

Of all the candidate companions (CCs), a total of 70 were cross-identified in the two data sets. Figure~\ref{fig:all_cc} presents the differential right ascension and declination of these 70 candidates between the two epochs, compared to the track expected for a stationary background object. Even though the astrometric error for some candidates is significant in the IRDIS data, the baseline of
more than 10 years is sufficient to unambiguously conclude that all these 70 candidates are background stars. Of the remaining candidates, 32 can be identified as background objects using their location in the color-magnitude diagram (CMD) in the IRDIS H2 and H3 filters. Figure~\ref{fig:cmd} shows all the candidates in a CMD compared to sequences of MLT objects from \citet{leggett2001}, \citet{Burgasser2014} and \citet{Schneider2015}. The candidates located near the zero-color and with $M_{H2} > 16$ cannot represent physical objects in orbit around \gsc\ and are therefore classified as background. Finally, the status of the remaining 7 candidates cannot be determined unambiguously. However, we note that 6 of them are located at separations larger than 5.5\arcsec, that
is, at projected physical separations larger than 400~au. The only candidate located at small angular separation (457 mas), which is also identified in the IFS field-of-view, presents a color $H_2 - H_3 = -0.07 \pm 0.1$  that is clearly offset at $\sim$2$\sigma$ with respect to the ML sequence (point identified by an arrow in Fig.~\ref{fig:cmd}). This offset is well aligned with the other candidates identified as background from proper motion (Fig.~\ref{fig:all_cc}). Moreover, its $J-H=0.78$ agrees with a background object. To further confirm its nature, we tested that this bright object would be visible in NIRC2 data \citep[$5\sigma$ contrast limit in $K_p$ is 8.9 mag at 0.5\arcsec,][]{galicher2016} if comoving, while it will be at about $50$ mas from the star if background and therefore not visible.  As a conclusion, all the candidates detected with projected physical separation below 400~au are identified as background stars.

\section{Modeling of the SPHERE/IRDIS data}

\subsection{Point source removal}
\label{sec:star_removal}
To reduce the effect of the field stars on the disk image, we removed the most contaminant field stars, replacing each star contribution in each raw individual frame of the data set by the background contribution, accounting for the parallactic rotation.  The stellar flux in is then replaced by the adjusted background inside a box of $17\times17$ pixel. These boxes rotated with the field of view during the sequence. After the ADI reduction steps, we removed residual artifacts from the bright CCs by replacing the $20\times20$ pixel box area centered on the CCs with the right orientation by a fitted background.
This new reduction removed the grooves created by the brightest field stars after application of the ADI and therefore better constraints on the disk properties are possible.

\subsection{Modeling strategy}
\label{sec:mod_strategy}

We used the same code as was used in \citet{Olofsson2016} and \citet{Feldt2017} and followed the same strategy. For the stellar parameters, we used an effective temperature of $3600$\,K and a distance of $73$\,pc. For the disk parameters, the free parameters were the following: the  inclination $i$, the position angle $\phi$, the reference radius $r_0$,  and the inner and outer slopes ($\alpha_{\mathrm{in}}$ and $\alpha_{\mathrm{out}}$, respectively) of the volumetric dust density distribution $n(r,z)$, which is defined as
\begin{equation}\label{eqn:nr}
n(r, z) \propto 
\left[\left(\frac{r}{r_0}\right)^{-2\alpha_{\mathrm{in}}} + \left(\frac{r}{r_0}\right)^{-2\alpha_{\mathrm{out}}}\right]^{-1/2} \times \mathrm{e}^{-z^2/2h^2},
\end{equation}
where $h = r \times \mathrm{tan}(\psi)$ and  $\psi$ is the opening  angle. To reduce the number of free parameters, we fixed $\psi = 0.04$ (\citealp{Thebault2009}). Overall, we did not resolve the disk in the vertical direction, and preliminary tests suggested that we cannot really constrain the value of $h$. For the dust properties, we assumed a grain size distribution that follows a differential power law d$n(s) \propto s^{-3.5}$d$s$, where $s$ is the grain size (\citealp{Dohnanyi1969}), between $s_{\mathrm{min}} = 0.01$\,$\mu$m and $s_{\mathrm{max}} = 1$\,mm. The optical constants are those for astro-silicates from \citet{Draine2003}, and the absorption and scattering efficiencies are computed using the Mie theory. To have a finer control on the shape of the scattering phase function $S_{11}$, we used the Henyey-Greenstein approximation rather than the Mie theory. This adds one free parameter $g,$ which governs the scattering efficiency as a function of the scattering angle $\theta$. Overall, because we modeled a monochromatic image and the scattered light contribution dominates the thermal emission in the near-infrared, the previous choices for the dust properties (e.g., $s_{\mathrm{min}}$ or $s_{\mathrm{max}}$) have little effect on the modeling results. The vertical distribution of the disk follows a Gaussian profile with a standard deviation $h  = 0.04 \times r$. 

To estimate the goodness of fit, it is first necessary to estimate the uncertainties. This is performed on the collapsed observations by measuring the standard deviation in concentric annuli with widths of $2$\,pixels. Because of the disk, we overestimate the proper uncertainties. 

For a given model and for each frame of the original datacube, we subtracted a scaled image of the model (the scaling factor being $10^f$, the last free parameter being $f$) derotated at the corresponding parallactic angle of that frame. We then performed the principal component analysis, removing the $\text{five}$ main components, and collapsed all the frames after derotating them of their parallactic angles. The goodness of fit is the sum of the squared residuals divided by the uncertainties. To speed up the modeling process, we cropped each frame of the datacube to have sizes of $300 \times 300$\,pixels. Additionally, we placed a numerical mask of $0.2\arcsec$ in radius.
\begin{figure*}
\centering
\includegraphics[width=\textwidth]{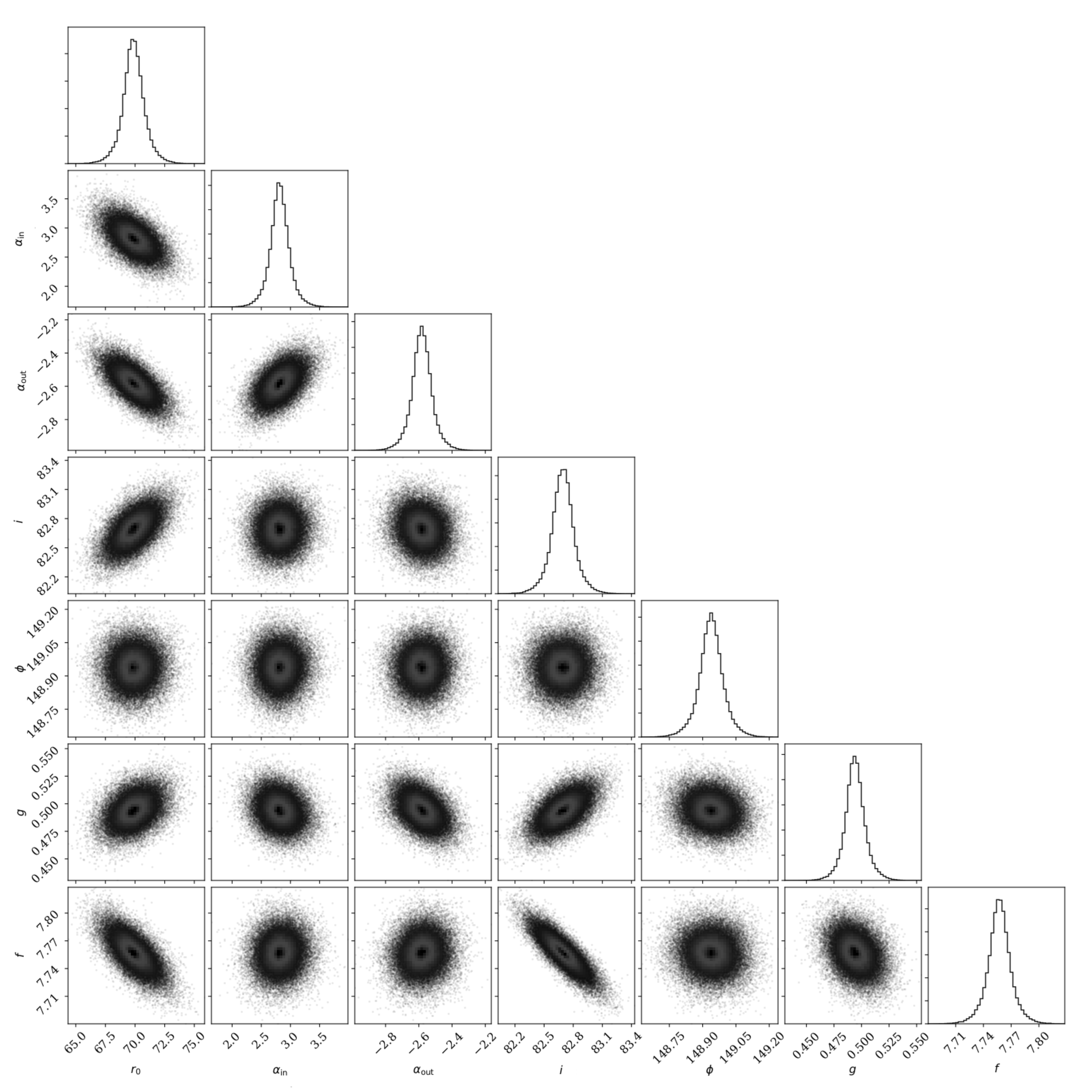}
\caption{Projected probability density distributions for the different parameters in the modeling as well as density plots.}
\label{fig:corner}
\end{figure*}
To summarize, we have $\text{seven}$ free parameters: $r_0$, $\alpha_{\mathrm{in}}$, $\alpha_{\mathrm{out}}$, $i$, $\phi$, $g$, and $f$, and we used an affine invariant ensemble sampler  to find the best solution (\citealp{Foreman-Mackey2012}). The Monte  Carlo Markov chain was composed of $100$ walkers, with an initial burn-in phase of $500$ steps, and it was run for additional $1000$ steps afterward. At the end of the modeling, the mean acceptance fraction was $0.498$ and the maximum autocorrelation time was $54$ steps (Fig.~\ref{fig:corner}).

\section{Mass limit estimation}
\label{sec:contrast}
\begin{figure}
\begin{tabular}{c}
\includegraphics[width=\columnwidth]{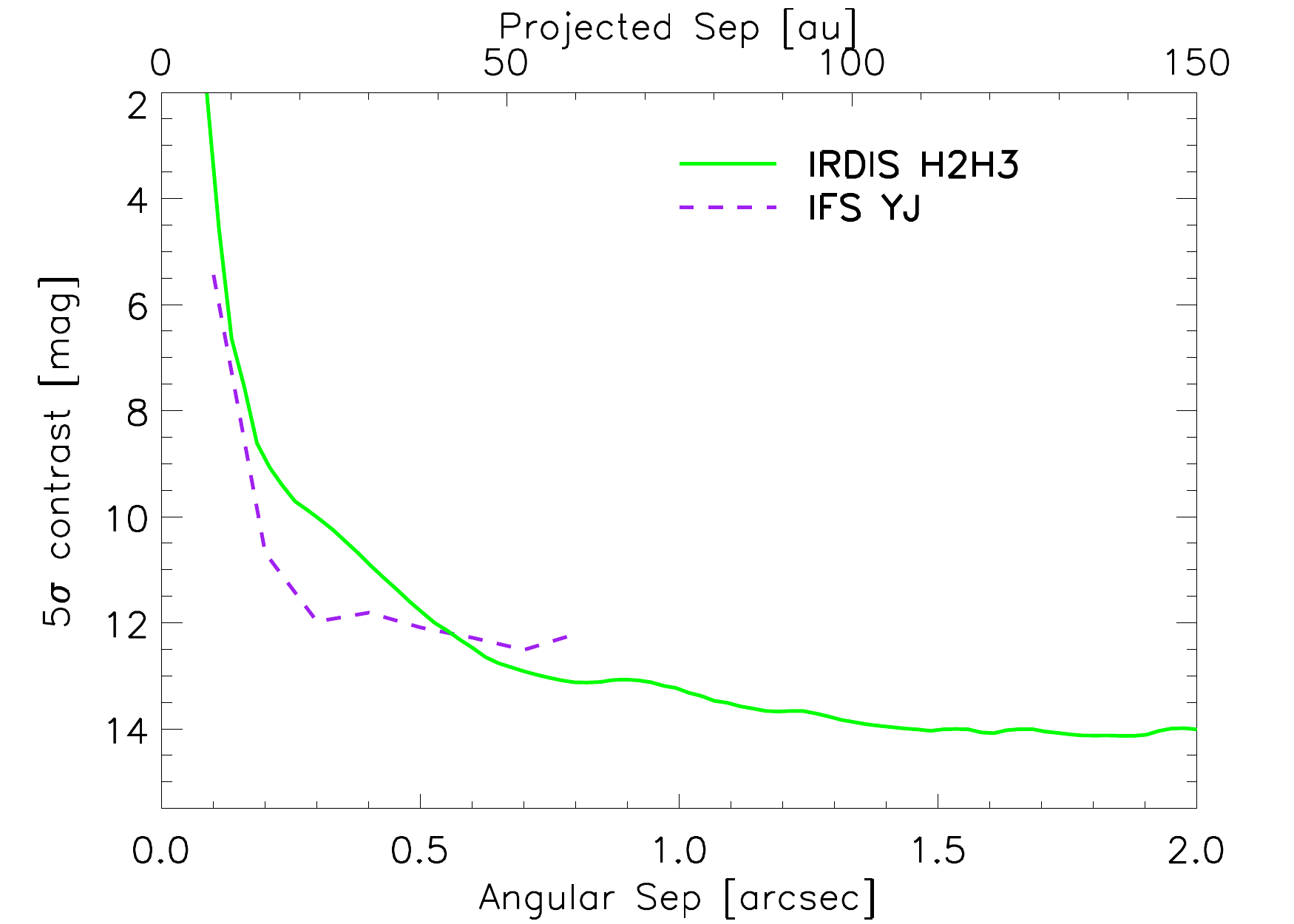}\\
\includegraphics[width=\columnwidth]{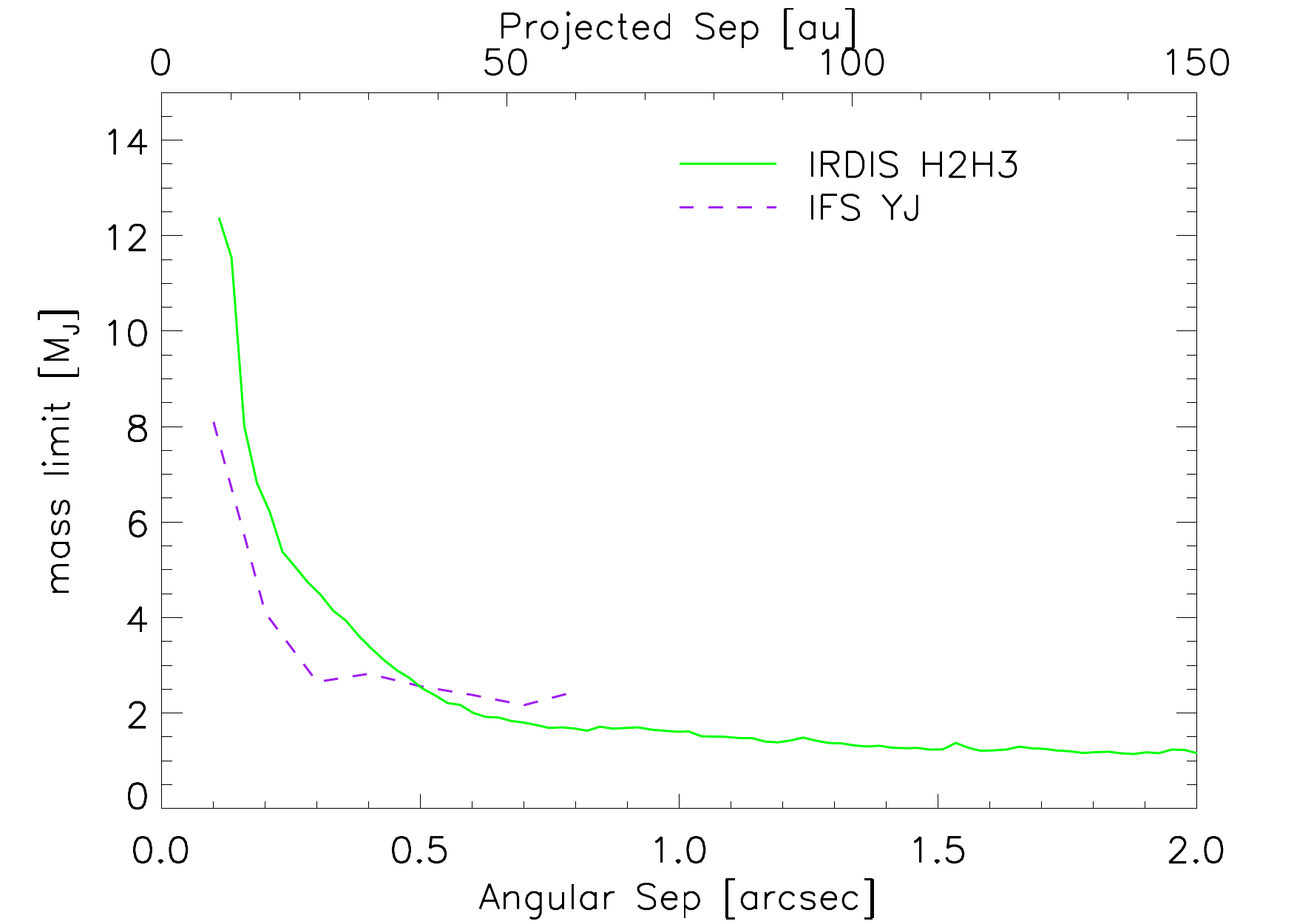} 
\end{tabular}
\caption{$5\sigma$ contrast curves and companion mass limits derived for IFS (dashed line) and for IRDIS H2H3 (solid line) using PCA reduction method and TLOCI reduction method, respectively, both after the brightest CCs subtraction as described in Appendix \ref{sec:star_removal}.}
\label{fig:contrast}
\end{figure}

The contrast and mass limit for unseen companions was determined for both IFS and IRDIS. Since the grooves related to the very bright stars after ADI were not negligible, we estimated the limits on the images after contaminant star removal (see Appendix~\ref{sec:star_removal} for details).
The azimuthally averaged $5\sigma$ contrast curves were estimated on the wavelength-collapsed monochromatic five-component PCA reduction for IFS and template locally optimized combination of images \citep[TLOCI,][]{marois2014} reduction for IRDIS, and they are shown in Fig.~\ref{fig:contrast} (upper panel). We then used the theoretical atmospheric models AMES-COND \citep{allard2003} to convert this limit into the unseen companion mass limit (Fig.~\ref{fig:contrast}, bottom panel).

\end{appendix}

\end{document}